# Dispatchable Region for Active Distribution Networks Using Approximate Second-Order Cone Relaxation

Zhigang Li, *Senior Member, CSEE*, Wenjing Huang, J. H. Zheng, *Member, IEEE,* and Q. H. Wu, *Fellow, IEEE*

*Abstract*—Uncertainty in distributed renewable generation threatens the security of power distribution systems. The concept of the dispatchable region was developed to assess the ability of power systems to accommodate renewable generation at a given operating point. Although DC and linearized AC power flow equations are typically used to model dispatchable regions for transmission systems, these equations are rarely suitable for distribution networks. To achieve a suitable trade-off between accuracy and efficiency, this paper proposes a dispatchable region formulation for distribution networks using tight convex relaxation. Second-order cone relaxation is adopted to reformulate the AC power flow equations, which are then approximated by a polyhedron to improve tractability. Further, an efficient adaptive constraint generation algorithm is employed to construct the proposed dispatchable region. Case studies on distribution systems of various scales validate the computational efficiency and accuracy of the proposed method.

## Nomenclature

### A. Indices and Sets

$(\cdot)_j$    Symbol associated with bus $j$.

$(\cdot)_{ij}$    Symbol associated with the line from bus $i$ to bus $j$.

$W(j)$    Set of renewable energy units connected to bus $j$.

### B. Input Parameters

$p_j^d$    Active load demand at bus $j$.

$p_j^g$    Active power generation at bus $j$.

$q_j^d$    Reactive load demand at bus $j$.

$q_j^g$    Reactive power generation at bus $j$.

$r_{ij}$    Resistance of the line from bus $i$ to bus $j$.

$\overline{S}_{ij}$    Maximum apparent power on the line from bus $i$ to bus $j$.

$\mu_n$    Constant related to the power factor of renewable generator $n$.

$w_n$    Actual active generation output of renewable generator $n$.

$w_n^e$    Predicted active generation output of renewable generator $n$.

$x_{ij}$    Reactance of the line from bus $i$ to bus $j$.

### C. Decision Variables

$\ell_{ij}$    Squared line current magnitude on the line from bus $i$ to bus $j$.

$p_j^c$    Rescheduling of active power generation at bus $j$.

$P_{ij}$    Active power flow on the line from bus $i$ to bus $j$.

$q_j^c$    Rescheduling of reactive power generation at bus $j$.

$Q_{ij}$    Reactive power flow on the line from bus $i$ to bus $j$.

$v_i$    Squared voltage magnitude at bus $i$.

$\Delta w_n$    Difference between the predicted output and actual output of renewable generator $n$.

$y$    Vector of all decision variables except $\Delta w$.

## I. Introduction

THE large-scale integration of renewable power generation (RPG) can markedly reduce fossil fuel consumption and help alleviate air pollution. On the other hand, RPG variability increases the risks of power balancing and transmission [1]. Sufficient power system flexibility is required to cope with these challenges. Reference [2] provides a comprehensive

This work was supported by the National Natural Science Foundation of China under Grant 52177086, the Guangdong Basic and Applied Basic Research Foundation under Grant 2019A1515011408, the Science and Technology Program of Guangzhou under Grant 201904010215, and the Talent Recruitment Project of Guangdong under Grant 2017GC010467.

The authors are with the School of Electric Power Engineering, South China University of Technology, Guangzhou 510641, China. (Corresponding author: J. H. Zheng, email address: zhengjh@scut.edu.cn)

review of power system flexibility with large-scale volatile RPG integration. Additionally, a geometric interpretation is provided through visualization to study the effects of various energy resources on grid-side flexibility [2].

Recently, the concept of the dispatchable region [3]-[5] was proposed to be used in place of the conventional probability distribution function to characterize the ability of a power system to accommodate fluctuating renewable energy. The dispatchable region is defined as the allowable range of uncertain nodal power injections that a power system can accommodate at a given operating point. Instead of providing a specific dispatch strategy, the dispatchable region is introduced to describe the largest ranges of RPG outputs that do not include infeasible operating points. It can be regarded as an index that quantifies the flexibility of a given dispatch strategy. Another relevant concept is the "security region", which is defined as a set of operation points that satisfy given operation and security constraints [6]. In fact, the dispatchable region, defined in the space of uncertain power injections, can be regarded as an extension of the security region.

The do-not-exceed (DNE) limit is the maximum inner box approximation of the dispatchable region for RPG, as described in [7]. A data-driven method was proposed in [8] to construct DNE limits to maximize RPG utilization. As an inner approximation of the dispatchable region, the box-type DNE limit appears to be conservative in general cases. As alternatives, polyhedral and elliptical sets have also been developed to describe the ranges of uncertain parameters [9], [10]. However, such DNE limits cannot accurately describe the real ability of a power system to accommodate power injection fluctuations. Thus, how best to formulate the exact dispatchable region remains an attractive research topic from both theoretical and practical perspectives.

The concepts of the dispatchable region and dispatchability were proposed for transmission networks based on DC power flows in [3] and [4]. Given an operating point, in real-time dispatch, all constraints are projected into the space of the uncertain parameters to yield the dispatchable region. Mathematically, the dispatchable region is represented as a polytope constructed by solving a series of mixed-integer linear programming (MILP) problems [4]. The shape of the dispatchable region can be tuned by means of energy-reserve co-dispatch [5]. Contingencies, which are commonly invoked in unit commitment [11], [12] and economic dispatch [13], [14], are not considered in dispatchable region analyses due to the limited decision time available for real-time dispatch. References [3]-[5] used DC power flow formulations without considering the voltage magnitude or reactive power, which may result in the inclusion of unsafe operating points in the dispatchable region. A dispatchable region model based on linearized AC power flows was proposed in [15], [16] that provides a high-quality approximation for high-voltage transmission networks.

Most of the aforementioned references have adopted pure DC power flow formulations, while reference [15] adopted a linearized AC power flow approach to formulate the dispatchable region. However, for a distribution network with high resistance-to-reactance (r/x) ratios of the transmission lines, linearization of the power flows and the consequent disregard of the inherent features of AC power flows may lead to infeasible power flow profiles. In this context, the nonlinearity of the power flows should be adequately addressed.

For a radial distribution network, the branch flow model (BFM), also known as *DistFlow* [17], is a commonly used description of the nonlinear AC power flows in which the phase angles are eliminated. However, the power flow equations in the BFM are nonlinear. The linearized *DistFlow* model [18] ignores the losses on the lines and assumes that the node voltages are approximately valued at 1 p.u., similar to the linearized AC power flow model devised in [15]. Although the linearized *DistFlow* model has the advantage of simplified equations, it does not provide a rigorous outer approximation of the *DistFlow* model; consequently, some feasible points may be lost, or additional infeasible regions may even be generated.

Convex approximation methods have been adopted in the literature to address the nonlinearity and nonconvexity of AC power flow equations. A convex inner approximation of the solvable region for power flow equations [19]-[20] usually leads to overly conservative results. A relaxed BFM was proposed in [21] and [22], in which the nonconvex constraints are relaxed into second-order cones (SOCs) in the optimal power flow (OPF) problem. Reference [23] analyzed the performance of semidefinite programming (SDP), chordal and SOC programming (SOCP) relaxations of the OPF problem. SOCP is the most computationally efficient approach, whereas SDP usually requires the most computational effort. The exactness of the relaxations depends on the network topology. For a radial network, the SOCP relaxation is the tightest and simplest among the three relaxation techniques. In fact, the SOC relaxation can be proven to be conditionally exact for a radial network. In [24], a SOC-based convex hull formulation of nonconvex quadratic equations was constructed, which was proven to be the tightest convex outer approximation of the nonconvex quadratic equations in the BFM [24]. To the best of our knowledge, however, little work has been done on the use of convex hulls to relax the AC power flows in dispatchable regions. In addition, solving for the dispatchable region is essentially a projection problem, and such problems are still challenging to handle even for convex hull models. Thus, it is still necessary to find an effective strategy for approximating the dispatchable region by taking advantage of convex hull relaxation.

To tackle these issues, this paper focuses on the dispatchable region for volatile RPG in a distribution network considering AC power flows. By combining SOC relaxation and polyhedral outer approximation, we develop an approach for constructing an approximate dispatchable region in a tractable fashion. The contributions of this paper are summarized as follows.

1) A dispatchable region model for distribution networks is devised based on the BFM. A tighter convex outer approximation of the dispatchable region is obtained by using the SOC-based convex hull of a quadratic equation. In contrast to the models presented in existing work, the proposed model captures all feasible operating points for nodal power injections in a power system, even with large differences in the r/x ratios of the lines.

2) A linear convex outer approximation of the dispatchable region is devised. A polyhedral outer approximation is used to linearize the SOC constraints. An adaptive constraint generation algorithm is exploited to construct the boundaries of the proposed dispatchable region.

The rest of this paper is organized as follows. The dispatchable region of a power distribution network is defined and formulated in Section II. The tight convex relaxation of the dispatchable region and the adaptive constraint generation (Ad-CG) algorithm are presented in Section III. Case studies are reported in Section IV. Conclusions are given in Section V.

## II. DEFINITION AND FORMULATION OF THE DISPATCHABLE REGION FOR A POWER DISTRIBUTION NETWORK

The volatility of RPG affects the power balance of real-time dispatch in a power system. In the current dispatch interval, the predispatch strategy for the traditional generators in the system is denoted by $\{p^g, q^g\}$, and the active outputs of the renewable generators may deviate from their predicted values $w^e$, such that their actual outputs are assumed to be $w = w^e + \Delta w$. Once the output deviations $\Delta w$ of the RPG units have been observed, a rescheduling strategy $\{p^c, q^c\}$ is implemented to restore the system to a secure operating state, and the outputs of the traditional generators are changed to $\{p^g + p^c, q^g + q^c\}$. Due to the limited ramping capability of generators and the physical constraints of the network, the system cannot accommodate arbitrarily large ranges of RPG fluctuation. Accordingly, the dispatchable region can be expressed in the following abstract form:

$$W(p^g, q^g, w^e) = \{\Delta w \mid \exists y : f(\Delta w, y) \leq 0\} \quad (1)$$

where $f(\Delta w, y) \leq 0$ represents the power flow equations and operating limits of the power system, $y$ represents the variables related to corrective actions that rebalance the power system, and $\Delta w$ is a parameter of $W$.

Here, the constraints $f(\Delta w, y) \leq 0$ of a single-phase or three-phase balanced distribution system include the following:

$$p_j^g + p_j^c - p_j^d = \sum_{k:j \to k} P_{jk} - (P_{ij} - r_{ij}\ell_{ij}) - \sum_{n \in W(j)} w_n \quad (2)$$

$$q_j^g + q_j^c - q_j^d = \sum_{k:j \to k} Q_{jk} - (Q_{ij} - x_{ij}\ell_{ij}) - \sum_{n \in W(j)} \mu_n w_n \quad (3)$$

$$v_j = v_i - 2(r_{ij}P_{ij} + x_{ij}Q_{ij}) + (r_{ij}^2 + x_{ij}^2)\ell_{ij} \quad (4)$$

$$P_{ij}^2 + Q_{ij}^2 = v_i \ell_{ij} \quad (5)$$

$$\underline{p}_j^g \leq p_j^g + p_j^c \leq \overline{p}_j^g, \quad \underline{q}_j^g \leq q_j^g + q_j^c \leq \overline{q}_j^g \quad (6)$$

$$w_n = w_n^e + \Delta w_n \leq \overline{w}_n \quad (7)$$

$$0 \leq |p_j^c| \leq Ramp_j \quad (8)$$

$$0 \leq \ell_{ij} \leq \overline{\ell}_{ij} \quad (9)$$

$$\underline{v}_j \leq v_j \leq \overline{v}_j \quad (10)$$

$$P_{ij}^2 + Q_{ij}^2 \leq \overline{S}_{ij}^2 \quad (11)$$

where the subscript $i$ denotes the bus index and the subscript $ij$ denotes the directed branch linking buses $i$ and $j$. The superscripts $g$, $c$, $d$, and $e$ represent the predispatch output of a traditional generator, the change in the rescheduled output of the traditional generator, the load on the bus, and the estimated RPG output, respectively.

Constraints (2)–(5) represent the BFM of the distribution network based on the *DistFlow* model [17]. Constraints (2) and (3) are the conditions for nodal active and reactive power balancing. Constraint (4) describes the voltage drop on each line. Constraint (5) describes the apparent power flow at the from-bus of each line. Constraint (6) gives the capacity restrictions of a traditional generator. Constraint (7) represents the active RPG output, where $\overline{w}_n$ is the maximum active output from renewable resources. Constraint (8) stipulates that the power regulation of a traditional generator is subject to the ramping capacity. Constraints (9), (10), and (11) represent the line current capacities, voltage magnitude constraints, and apparent power flow limits of the distribution lines, respectively.

The load of the current system and the predicted active power output $w^e$ from RPG are assumed to be predefined by forecasting algorithms. On this basis, the current predispatch strategy $\{p^g, q^g\}$ can be calculated. $y$ includes all variables except $p^g$, $q^g$ and $\Delta w$ in constraints (2)–(11).

## III. TIGHT CONVEX RELAXATION OF THE DISPATCHABLE REGION AND ITS CONSTRUCTION

The BFM presented by (2)–(5) is nonconvex and nonlinear, resulting in a nonconvex dispatchable region. In this section, to make the construction of the dispatchable region tractable, an SOC-based convex hull of the original dispatchable region is developed to provide a tight outer approximation of the original dispatchable region. The SOC-based convex hull is an approximate SOC relaxation. This convex hull is further relaxed into a polyhedral dispatchable region, which can be readily constructed using the Ad-CG algorithm.

### A. Polyhedral Outer Approximation Based on Convex Hull Relaxation

The SOC-based convex hull (CH) approach, analytically proven and geometrically verified in [24], is adopted to outer approximate the dispatchable region. Let $\Omega_0$ denote the feasible set described by the quadratic equality (5) and (9)–(11) for each branch $(i,j)$, and let $\Omega_1$ be the CH of $\Omega_0$, which are formulated as follows:

$$\Omega_0 = \{\eta_{ij} \mid (5) \text{ and } (9)-(11) \text{ hold}\} \quad (12)$$

$$\Omega_1 = \text{CH}(\Omega_0) = \left\{ \eta_{ij} \left| \begin{array}{l} \|A\eta_{ij}\|_2 - s^T \eta_{ij} \leq 0 \\ c_{ij}^T \eta_{ij} - d_{ij} \leq 0 \\ \eta_{ij} \in (9)-(11) \end{array} \right. \right\} \quad (13)$$

where $\boldsymbol{\eta}_{ij} = [P_{ij}, Q_{ij}, \ell_{ij}, v_i]^T$, $A = \text{diag}([\sqrt{2}, \sqrt{2}, 1, 1]^T)$ and $\boldsymbol{s} = [0, 0, 1, 1]^T$. The superscript $T$ is the transpose symbol. $\|A\boldsymbol{\eta}_{ij}\|_2 - \boldsymbol{s}^T\boldsymbol{\eta}_{ij} \leq 0$ in $\Omega_1$ represents the SOC of quadratic equality (5) in the following form:

$$P_{ij}^2 + Q_{ij}^2 \leq v_i \ell_{ij}. \tag{14}$$

which is the convex relaxation of (5).

The values of the coefficient vectors in

$$\boldsymbol{c}_{ij}^T \boldsymbol{x}_{ij} - d_{ij} \leq 0 \tag{15}$$

depend on specific constraints. When $\overline{S}_{ij}^2 \leq \overline{\ell}_{ij}\underline{v}_i$, the branch flow is bounded by (11), with (9) being redundant; then, we have $\boldsymbol{c}_{ij} = [0, 0, \underline{v}_i\overline{v}_i, \overline{S}_{ij}^2]^T$ and $d_{ij} = (\underline{v}_i + \overline{v}_i)\overline{S}_{ij}^2$. When $\overline{S}_{ij}^2 \geq \overline{\ell}_{ij}\underline{v}_i$, we have $\boldsymbol{c}_{ij} = [0, 0, \overline{v}_i, \overline{\ell}_{ij}]^T$ and $d_{ij} = \overline{\ell}_{ij}\overline{v}_i + \overline{S}_{ij}^2$. When $\overline{S}_{ij}^2 \geq \overline{\ell}_{ij}\overline{v}_i$, (11) is redundant. Constraint (11) usually dominates over (9) for long transmission lines, and the converse may occur for distribution feeders.

The relation between $\Omega_0$ and $\Omega_1$ is shown in Fig. 1. The extreme points of the CH belong to the original nonconvex set.

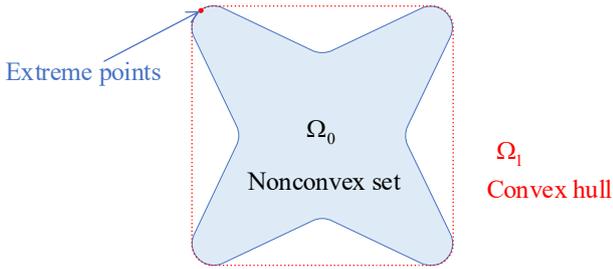

Fig. 1. Illustration of the advantages of the CH.

The sufficient conditions for the exact convex relaxation are summarized in [25]. Since CH relaxation introduces an SOC, the boundary of the obtained dispatchable region is still nonlinear. Thus, to construct a dispatchable region with linear boundaries, the technique in [26] is used to generate a polyhedral outer approximation of the dispatchable region. The rotated SOC (14) in $\mathbb{R}^4$ is split into a combination of two standard SOCs in $\mathbb{R}^3$ as follows:

$$m_{ij} \geq \sqrt{P_{ij}^2 + Q_{ij}^2} \tag{16}$$

$$m'_{ij} \geq \sqrt{P_{ij}^{\prime 2} + Q_{ij}^{\prime 2}} \tag{17}$$

$$Q'_{ij} = m_{ij} = \frac{v_i - \ell_{ij}}{2} \tag{18}$$

$$m'_{ij} = \frac{v_i + \ell_{ij}}{2} \tag{19}$$

where $m_{ij}, m'_{ij}, P'_{ij}$ and $Q'_{ij}$ are auxiliary variables. Then, auxiliary variables $P_{ij,n}$ and $Q_{ij,n} (n=1,\cdots,k)$ are introduced to obtain a polyhedral outer approximation of the standard SOC (16) in $\mathbb{R}^3$ in the following form:

$$P_{ij,n+1} - P_{ij,n}\cos\frac{\pi}{2^n} - Q_{ij,n}\sin\frac{\pi}{2^n} = 0, \; n = 0,\cdots,k-1 \tag{20}$$

$$Q_{ij,n+1} - Q_{ij,n}\cos\frac{\pi}{2^n} + P_{ij,n}\sin\frac{\pi}{2^n} \geq 0, \; n = 0,\cdots,k-1 \tag{21}$$

$$Q_{ij,n+1} + Q_{ij,n}\cos\frac{\pi}{2^n} - P_{ij,n}\sin\frac{\pi}{2^n} \geq 0, \; n = 0,\cdots,k-1 \tag{22}$$

$$P_{ij,k}\cos\frac{\pi}{2^k} + Q_{ij,k}\sin\frac{\pi}{2^k} - m_{ij} = 0 \tag{23}$$

Note that equations (20) and (23) can eliminate the auxiliary variables $P_{ij,n}$ ($n = 1,\cdots,k$) and $Q_{ij,k}$ in inequalities (21) and (22). $k$ is the number of auxiliary variables introduced by each of $P_{ij}$ and $Q_{ij}$ separately. In this way, we obtain $2k$ inequalities with $k+2$ variables that are expressed as

$$g(m_{ij}, P_{ij}, Q_{ij}, Q_{ij,1}, \cdots, Q_{ij,k-1}) \leq 0 \tag{24}$$

Similarly, the standard SOC (17) in $\mathbb{R}^3$ can also be approximated as a set of $2k$ inequalities with $k+2$ variables:

$$g(m'_{ij}, P'_{ij}, Q'_{ij}, Q'_{ij,1}, \cdots, Q'_{ij,k-1}) \leq 0 \tag{25}$$

The set of inequalities in (24) constructs a high-dimensional polyhedron space, whose projection on the hyperplane of the SOC given in (16) is a tight outer approximation of the SOC. The relaxation accuracy $\varepsilon_1 = \cos^{-1}(\pi/2^k) - 1$ satisfies [27]

$$1 + \varepsilon_1 \geq \frac{\sqrt{P_{ij}^2 + Q_{ij}^2}}{m_{ij}} \tag{26}$$

Suppose that the relaxation accuracy of the polyhedral approximation (25) is $\varepsilon_2$; then, the overall approximation accuracy $\varepsilon$ of the rotated SOC (14) is

$$\varepsilon = (1+\varepsilon_1)(1+\varepsilon_2) - 1 \tag{27}$$

The larger $k$ is, the higher the overall approximation accuracy.

*B. Linear Relaxation of Quadratic Inequality Constraints*

A quadratic inequality constraint of the form given in (11) is a type of power circle constraint, which can also be approximated via polyhedral constraints. Notably, constraint (11) tends to be redundant in distribution networks.

For a circle with a radius of $RR$, the area inside the circle $Rx^2 + Ry^2 \leq RR^2$ can be approximated as illustrated in Fig. 2 by introducing $t$ circumscribed squares to generate $4t$ inequality constraints:

$$-RR \leq Rx \leq RR, \; -RR \leq Ry \leq RR \tag{28}$$

$$-\frac{RR}{\sin\theta} \leq Ry - \cot\theta \leq \frac{RR}{\sin\theta} \tag{29}$$

$$-\frac{RR}{\cos\theta} \leq Ry + \tan\theta \leq \frac{RR}{\cos\theta} \tag{30}$$

$$\theta = \frac{z\pi}{2t}, \; t > 1, \; z = 1,\cdots,t-1 \tag{31}$$

Therefore, each distribution line power flow limit described by (11) can be approximated as a series of linear inequalities:

$$h(P_{ij}, Q_{ij}) \leq 0 \tag{32}$$

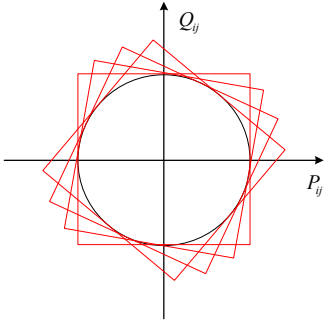

Fig. 2. Polyhedral outer approximation of a power circle.

*C. Efficient Ad-CG Algorithm for Solving for the Dispatchable Region*

All the polyhedral approximations given above are strictly convex relaxations. The solution space of the dispatchable region formed by these polyhedrons contains all feasible operating points.

In summary, the dispatchable region after tight convex relaxation can be expressed as

$$W^{\text{TCR}} = \{\Delta w \mid \exists y : By + C\Delta w \le b\} \quad (33)$$

where $y$ includes all power flow variables and auxiliary variables except $p^g$, $q^g$ and $\Delta w$. The linear constraints in (33) include (2)–(4), (6)–(10), (15), (24), (25), and (32). The procedure for constructing the proposed dispatchable region is illustrated in Fig. 3.

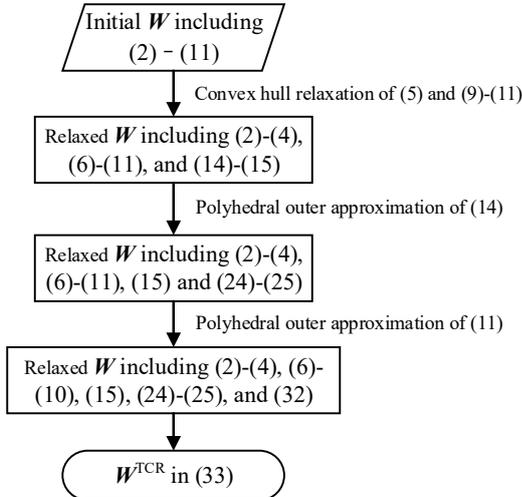

Fig. 3. Flowchart of the procedure for constructing the proposed dispatchable region.

Here, $W^{\text{TCR}}$ is a high-dimensional polyhedron. Solving for $W^{\text{TCR}}$ involves projecting a high-dimensional polyhedron onto the hyperplane of $\Delta w$. By taking the dual form of the linear constraints related to $y$, (33) is found to be equivalent to the following polyhedron [4]:

$$W^{\text{TCR}} = \{\Delta w \mid u^T C \Delta w \ge u^T b, \forall u \in \text{vert}(U)\} \quad (34)$$

where $U = \{u \mid B^T u = 0, -1 \le u \le 0\}$. The polytope $U$ has a finite number of vertices, and its coefficient matrix $B$ is related only to the structure of the power system. $U$ is independent of the current operating conditions and is fixed for every given system. Therefore, the specific expression of (34) can be obtained by enumerating the vertices of $U$.

In small-scale power systems, the vertices of $U$ can be calculated offline in advance. However, vertex enumeration produces a large number of redundant constraints and is computationally intensive. Enumerating all the vertices of $U$ is not feasible for a practical-scale system.

The Ad-CG algorithm [4] is employed to identify the binding vertices in $U$ and adaptively generate the boundary of $W^{\text{TCR}}$ without seeking any boundary points. In this algorithm, a sufficiently large initial set $W^B \supseteq W^{\text{TCR}}$ is first generated. Then, the algorithm finds all points that satisfy $\Delta w^* \notin W^{\text{TCR}}$ and creates corresponding cut plane constraints to iteratively join $W^B$ until $W^B \subseteq W^{\text{TCR}}$, which means $W^B = W^{\text{TCR}}$. The details of the Ad-CG algorithm are summarized as follows.

**Ad-CG Algorithm for Calculating the Dispatchable Region**

1: Choose a tolerance $\delta > 0$ and a sufficiently large set $W^B = \{\Delta w \mid H\Delta w \ge h\}$. Set $R = +\infty$.

2: While $R > \delta$, solve the following MILP problem:

$$R = \max \; u^T b + \xi^T h$$
$$\text{s.t.} \; C^T u + H^T \xi = 0$$
$$-M\theta \le h - H\Delta w \le 0 \quad (35)$$
$$-M(1-\theta) \le \xi \le 0$$
$$u \in U, \theta \in \{0,1\}^N$$

The optimal solution is $(u^*, \Delta w^*)$, and the optimal value is $R$. The decision variables of the MILP problem in step 2 include the RPG forecast errors $\Delta w$ and the dual variables $u$ and $\xi$. The number of inequality constraints in $W^B$ is $N$. The vector $\theta$ includes $N$ binary variables, and $M$ is a sufficiently large constant.

3: If $R \le \delta$, terminate, and report $W^{\text{TCR}} = W^B$; otherwise, add the following constraint into $W^B$:

$$(u^*)^T C \Delta w \ge (u^*)^T b \quad (36)$$

Then, update the matrix $H$ and the vector $h$ in $W^B$ and return to step 2.

## IV. CASE STUDY

The performance of the proposed model was tested using two numerical examples. All numerical experiments were conducted on a desktop computer with an Intel i5-8500 CPU and 8 GB of memory. All MILP problems were solved using Gurobi 9.10. The voltage level of the distribution networks in both test cases is 10 kV. The power factor of each RPG unit is 0.95 [28].

*A. Case 1: Performance When Constructing the Dispatchable Region and Impact of the Network Parameters*

A modified IEEE 33-bus distribution system [29] is adopted in this case, as shown in Fig. 4. Two RPG units, W12 and W26, are connected at buses #12 and #26, with capacities of 0.5 MW and 0.9 MW, respectively. The tolerance $\delta$ is set to $10^{-4}$. The up/down ramping limit of each traditional generator is set to 25% of its capacity.

*1) Construction of the Dispatchable Region*

To verify the effectiveness of the proposed method, we compare the following three types of dispatchable regions:

a) The exact dispatchable region ($W$), which is described by the feasible points of RPG outputs during power system operation. By using the MATPOWER program [30] to sample and calculate in the parameter space $\Delta w$, these feasible points of optimal AC power flows can be obtained.

b) The linearized approximation of the dispatchable region ($W^{\text{LA}}$) proposed in [15].

c) The tight convex relaxation of the dispatchable region ($W^{\text{TCR}}$) given by (34).

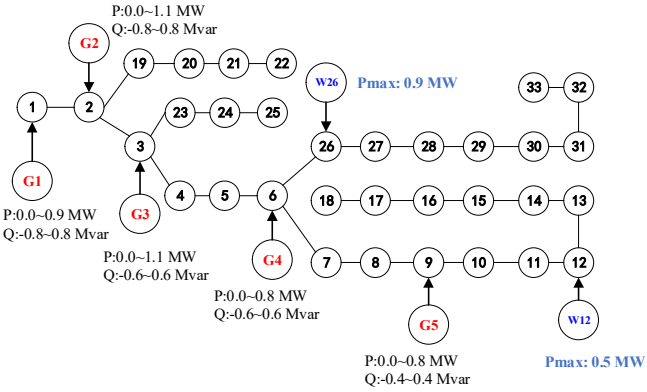

Fig. 4. Modified IEEE 33-bus distribution system structure and generator parameters

These dispatchable regions are shown in Fig. 5. The exact $W$ is obtained through sample approximation with a resolution of 0.03 MW. The proposed $W^{\text{TCR}}$ covers $W$ completely, and it also contains a small region of infeasible points. $W^{\text{LA}}$ also covers $W$ but contains a larger infeasible region. Thus, the polyhedron $W^{\text{TCR}}$ is validated to be a tighter convex relaxation of $W$ than $W^{\text{LA}}$.

The effective area percentage is defined to measure the similarity between two dispatchable regions as follows:

$$EP(W^*) = \frac{V(W)}{V(W^*)} \times 100\%, \quad (37)$$

where $EP(W^*)$ is the effective percentage of dispatchable region $W^*$ and $V(\cdot)$ is the volume of the corresponding dispatchable region. Because $W^{\text{TCR}}$ is a relaxation of the exact $W$, the value of $EP(W^{\text{TCR}})$ is less than one. We use $EP(W^{\text{TCR}})$ to indirectly evaluate the error of the proposed model, i.e., the closer to 1 $EP(W^{\text{TCR}})$ is, the smaller the error is.

The $EP$ for each dispatchable region is listed in the legend of Fig. 5. Trivially, $EP(W)$ for the dispatchable region obtained via the sampling method is 100%. $EP(W^{\text{TCR}})$ reaches 96.21%, indicating that $W^{\text{TCR}}$ is a better approximation than $W^{\text{LA}}$, which has an effective percentage of only 72.58%. In contrast to the exact $W$, there are still some infeasible points in $W^{\text{TCR}}$. For example, (0.5, 0.48) is an infeasible point in $W^{\text{TCR}}$ that violates the ramping limits of G4. In this scenario, the active power injection at bus #6 reaches the lower bound.

The convergence performance when constructing $W$, $W^{\text{LA}}$ and $W^{\text{TCR}}$ is shown in Fig. 6 and TABLE I. For this 33-bus distribution system, the construction of $W^{\text{LA}}$ needs only 17 iterations to converge because it involves fewer constraints and variables. The construction of $W^{\text{TCR}}$ involves more auxiliary variables and outer approximation constraints and consequently requires 36 iterations in total.

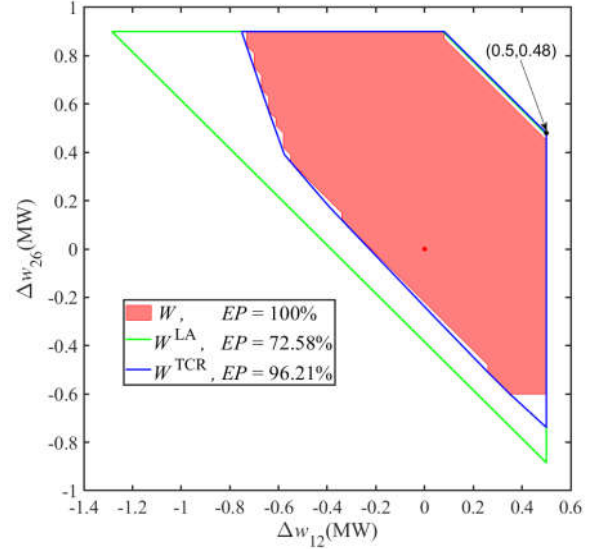

Fig. 5. Dispatchable regions obtained using the three methods.

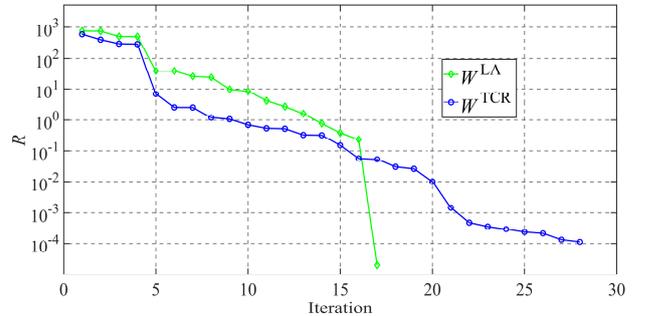

Fig. 6. Iterative process of calculating $W^{\text{LA}}$ and $W^{\text{TCR}}$.

TABLE I
PERFORMANCE OF DIFFERENT METHODS OF CONSTRUCTING DISPATCHABLE REGIONS

| Method | Computation time (s) | Number of iterations |
|---|---|---|
| $W$ | 140.25 | 2500 |
| $W^{\text{LA}}$ | 0.95 | 17 |
| $W^{\text{TCR}}$ | 7.34 | 36 |

*2) Effect of Polyhedral Approximation on the Accuracy of the Dispatchable Region*

In the process of constructing $W^{\text{TCR}}$, the parameter $k$ determines the scale of the introduced auxiliary variables and outer approximation constraints, which directly affects the computational efficiency. To balance accuracy and computational complexity, an appropriate value of $k$ should be chosen. Fig. 7 shows the $EP$, computation time and $\varepsilon$ of $W^{\text{TCR}}$ under different values of $k$, where $\varepsilon$ is the overall approximate accuracy of the rotated SOC (14). When $k$ is equal to 5, a sufficient-quality $W^{\text{TCR}}$ can already be obtained. Continuing to increase $k$ can result in a small further improvement in accuracy but leads to a large unnecessary computational burden. Accordingly, $k$ is set to 6 in this paper.

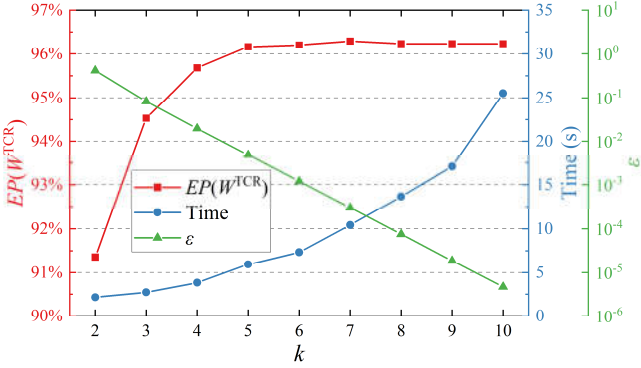

Fig. 7. Accuracy and computation time of $W^{\text{TCR}}$ under different $k$.

*3) Dispatchable Regions with a Fixed r/x Ratio*

In a distribution network, the r/x ratio of each line is a key parameter that affects the power supply capacity of the power system. In this subsection, we study dispatchable regions under the same r/x ratio but with different values of resistance and reactance. Four scenarios are considered with the same r/x ratio and various values of r and x, i.e., changes of +50%, +25%, -25% and -50% to their original values. The dispatchable regions are shown in Fig. 8.

As the feeder resistance and reactance increase with the same r/x ratio, the system transmission loss and the voltage drop also increase, affecting the available transmission capacity of the system. Consequently, the dispatchable region shrinks. When the resistance and reactance are increased by 50%, the dispatchable region does not include the origin, which implies that the generation output of the conventional units alone can no longer meet the load demand at this time.

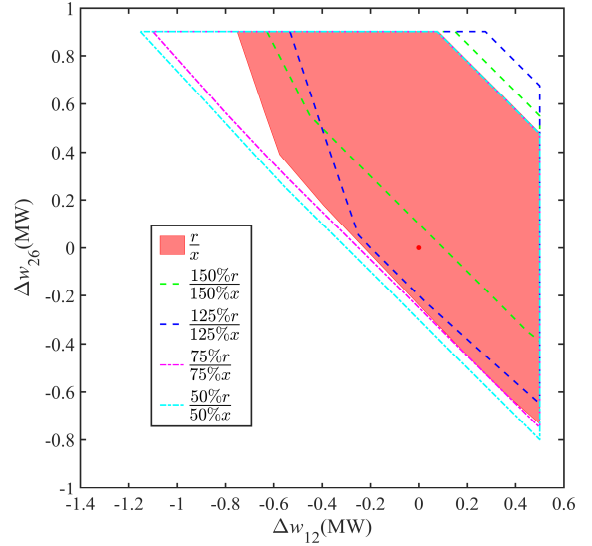

Fig. 8. Dispatchable regions with the same r/x ratio.

*B. Case 2: Computational Performance in Larger-Scale Systems with Multiple RPG Units*

*1) Three-Dimensional Dispatchable Region*

In this case, we include three RPG units in the 33-bus, 69-bus and 141-bus distribution systems. The sum of the capacity of the RPG units is set to 50% of the original generation capacity of all units in the corresponding system. The tolerance $\delta$ is set to $10^{-2}$.

The computational performance when constructing $W^{\text{TCR}}$ for the different distribution systems is shown in TABLE II. For the 33-bus system, the proposed method converges in 27 iterations, consuming approximately 15 seconds. The numbers of variables and constraints in (34) increase linearly with the number of nodes. Therefore, the computation times for the larger-scale systems are significantly longer.

The three-dimensional dispatchable region of the 141-bus system is shown in Fig. 9. The dark regions in $W$ represent the concave parts of the dispatchable region. $W^{\text{TCR}}$ can still cover the entirety of the exact $W$. Since constraint (7) imposes upper limits on the RPG outputs, the error of the proposed dispatchable region mainly depends on the lower limits of the RPG outputs. In these systems with three RPG units, $EP(W^{\text{TCR}})$ does not change significantly with the system scale. The $EP$ of the proposed method in a distribution system with three RPG units is greater than 82%.

TABLE II
COMPUTATIONAL PERFORMANCE FOR DIFFERENT SYSTEMS WITH THREE RPG UNITS

| System | Time (s) | Iterations | $EP(W^{\text{TCR}})$ | Variables | Constraints |
|---|---|---|---|---|---|
| 33-bus | 15.38 | 27 | 82.14% | 643 | 2157 |
| 69-bus | 25.25 | 19 | 82.56% | 1363 | 4569 |
| 141-bus | 135.76 | 28 | 83.91% | 2803 | 9393 |

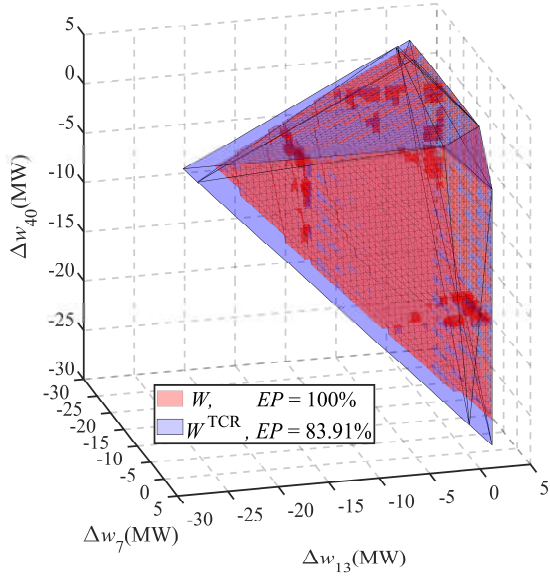

Fig. 9. Dispatchable regions of the IEEE 141-bus system with three RPG units

*2) IEEE 141-Bus Distribution System with Multiple RPG Units*

The performance when calculating the dispatchable regions for the 141-bus system with various numbers of RPG units is summarized in TABLE III.

$W^{LA}$ is the least computationally intensive to obtain. However, a larger infeasible region is included in the high-dimensional space, which tends to provide misleading information to the system operator. $W^{TCR}$ costs more computation time than $W^{LA}$, but it is still satisfactory for offline applications. When there are multiple RPG units in the power system, $W^{TCR}$ can provide useful security boundary information for day-ahead scheduling.

TABLE III
PERFORMANCE OF DIFFERENT METHODS FOR THE IEEE 141-BUS SYSTEM WITH DIFFERENT NUMBERS OF RPG UNITS

| Number of RPG units | Method | Time (s) | Iterations | $EP(\cdot)$ |
|---|---|---|---|---|
| 1 | $W$ | 2.91 | 50 | 100% |
| | $W^{LA}$ | 0.69 | 7 | 81.56% |
| | $W^{TCR}$ | 16.65 | 13 | 94.87% |
| 2 | $W$ | 147.00 | $50^2$ | 100% |
| | $W^{LA}$ | 0.76 | 11 | 70.22% |
| | $W^{TCR}$ | 60.50 | 22 | 88.24% |
| 3 | $W$ | 5092.26 | $50^3$ | 100% |
| | $W^{LA}$ | 0.80 | 12 | 62.37% |
| | $W^{TCR}$ | 135.76 | 28 | 83.91% |
| 4 | $W$ | 16103.14 | $25^4$ | 100% |
| | $W^{LA}$ | 17.36 | 29 | 49.77% |
| | $W^{TCR}$ | 386.05 | 41 | 69.81% |
| 5 | $W$ | 410678.91 | $25^5$ | 100% |
| | $W^{LA}$ | 36.66 | 59 | 23.64% |
| | $W^{TCR}$ | 678.88 | 48 | 51.13% |

This case demonstrates that the proposed dispatchable region model based on tight convex relaxation provides an effective analytical tool for studying the ability of power systems to accommodate uncertain RPG. The proposed method costs less computational time than $W$ while offering a higher effective percentage than $W^{LA}$, although its computational efficiency could still be further improved. A possible way to improve the accuracy would be to divide the RPG units into several groups and compute the dispatchable region for each of these groups in parallel. This could be an interesting direction for future work.

*3) Numerical Analysis of Relaxation Exactness*

Generally, medium- and high-voltage distribution networks are equipped with adequate reactive power compensation devices, so the reserve capacity of reactive power is sufficient. The voltage magnitudes in this case do not reach their lower or upper bounds, and the sufficient conditions for exact SOC relaxation presented in [21] are satisfied. The sum of the maximum RPG outputs in the distribution network is set to be smaller than the sum of the loads. The system can accommodate the maximum output of a single RPG unit but not the maximum outputs of all RPG units at the same time, as shown in Fig. 10. Therefore, the boundaries in the increasing directions of each RPG output are exact. When the RPG outputs decrease, the conventional generators are required to fill the active power mismatch, in which case the ramping limits are active and the sufficient conditions for exact SOC relaxation presented in [21] are not satisfied.

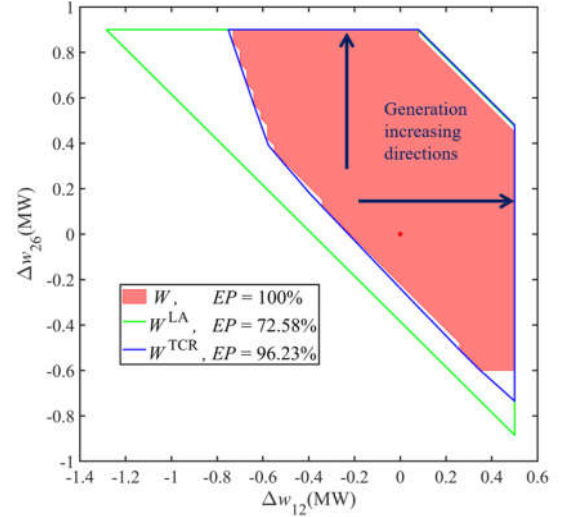

(a) Dispatchable region of the 33-bus system with increasing generation directions.

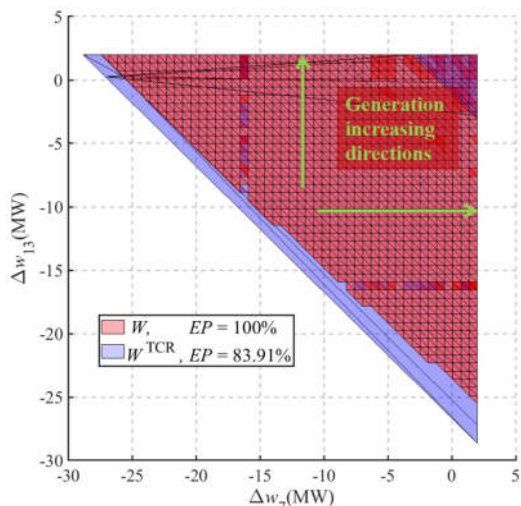

(b) A projection of the dispatchable regions of the 141-bus system with increasing generation directions.

Fig. 10. Two illustrative examples to analyze the exactness of relaxation.

## V. Conclusion

This paper proposes a dispatchable region model for a distribution network using the CH of the BFM for power flows. The proposed model is a tight convex outer approximation of the actual dispatchable region, which accurately describes the ability of the distribution system to accommodate fluctuations in renewable energy. A polyhedral projection method is introduced to obtain a tight linear convex outer approximation of the SOC constraint in the CH. The efficient Ad-CG algorithm is used to calculate the boundaries of our dispatchable region.

Numerical experiments conducted on various distribution systems of different scales are reported. The dispatchable regions under different network parameters show that the proposed method can be well adapted to the distribution system even with a high r/x ratio. The computation times under different numbers of RPG units demonstrate that the proposed method meets the computational requirements for practical use. The proposed dispatchable region is validated to be a high-quality approximation of the exact dispatchable region, allowing system operators to use this dispatchable region to analyze the security margin of the power grid.

Some interesting research directions remain. For example, by combining historical RPG data to determine the critical boundaries of the dispatchable region [16], the relationship between the critical boundaries and the physical constraints of the system could be studied in a quantitative way. This would be of great significance to provide suggestions for repairing the weak points of the system through the critical boundaries. Another interesting topic is to use parallel computing methods to address issues related to dispatchable region calculation, which is a focus of our ongoing studies.

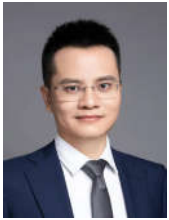

**Zhigang Li** received the Ph.D. degree in electrical engineering from Tsinghua University, Beijing, China, in 2016. He is currently an Associate Professor with the School of Electric Power Engineering, South China University of Technology, Guangzhou, China. He was a Visiting Scholar with the Illinois Institute of Technology, Chicago, IL, USA, and Argonne National Laboratory, Argonne, IL, USA. He is a Senior Member of IEEE, and a Senior Member of CSEE. He serves as an Associate Editor for the CSEE Journal of Power and Energy Systems. His research interests include smart grid operation and control, integrated energy systems, optimization theory and its application in energy systems.

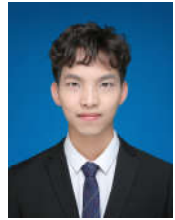

**Wenjing Huang** received the B.E. degree in electrical engineering from South China University of Technology, Guangzhou, China, in 2019. He is currently pursuing the M.E. degree at South China University of Technology. His research interest is the dispatchable region of the integrated energy system.

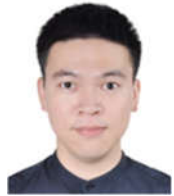

**J. H. Zheng** obtained his B.E. degree in Electrical Engineering from Huazhong University of Science and Technology, Wuhan, China, in 2012, and his Ph.D degree at the same area in South China University of Technology (SCUT), Guangzhou, China in 2017. He is currently a Lecturer in SCUT. His research interests include the area of optimization algorithms, decision making methods and their applications on integrated energy systems. He has authored or co-authored more than 40 peer-reviewed SCI journal papers.

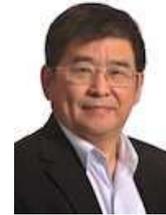

**Q. H. Wu** received the M.Sc. degree in electrical engineering from the Huazhong University of Science and Technology, Wuhan, China, in 1981, and the Ph.D. degree in electrical engineering from the Queen's University of Belfast (QUB), Belfast, U.K., in 1987. He is currently a Distinguished Professor with the School of Electric Power Engineering, South China University of Technology, Guangzhou, China, where he is the Director of the Energy Research Institute. He has authored and coauthored over 440 technical publications, including 220 journal papers, 20 book chapters, and 3 research monographs published by Springer. His research interests include nonlinear adaptive control, mathematical morphology, evolutionary computation, power quality, and power system control and operation. He is a fellow of IET and InstMC, and a Chartered Engineer.